\documentclass[twocolumn]{aastex631}

\shorttitle{JWST spectroscopy of planetary-mass objects}
\shortauthors{Damian et al.}
\usepackage{needspace}

\begin{document}

\title{Spectroscopy of Free-Floating Planetary-Mass Objects and their disks with JWST}

\author[0000-0002-2234-4678]{Belinda Damian}
\affiliation{SUPA, School of Physics \& Astronomy, University of St Andrews, North Haugh, St Andrews, KY16 9SS, UK}
\email{bd64@st-andrews.ac.uk}

\author[0000-0001-8993-5053]{Aleks Scholz}
\affiliation{SUPA, School of Physics \& Astronomy, University of St Andrews, North Haugh, St Andrews, KY16 9SS, UK}

\author[0000-0001-5349-6853]{Ray Jayawardhana}
\affiliation{Department of Physics \& Astronomy, Johns Hopkins University,  Baltimore, MD, 21218, USA}

\author[0000-0002-4945-9483]{V. Almendros-Abad}
\affiliation{Istituto Nazionale di Astrofisica (INAF) - Osservatorio Astronomico di Palermo, Piazza del Parlamento 1, 90134, Palermo, Italy}

\author[0000-0001-6362-0571]{Laura Flagg}
\affiliation{Department of Physics \& Astronomy, Johns Hopkins University,  Baltimore, MD, 21218, USA}

\author[0000-0002-7989-2595]{Koraljka Mu\v{z}i\'c}
\affiliation{Instituto de Astrof\'{i}sica e Ci\^{e}ncias do Espaço, Faculdade de Ci\^{e}ncias, Universidade de Lisboa, Ed. C8, Campo Grande, 1749-016 Lisbon, Portugal}

\author{Antonella Natta}
\affiliation{School of Cosmic Physics, Dublin Institute for Advanced Studies, 31 Fitzwilliam Place, Dublin 2, Ireland}

\author[0000-0001-8764-1780]{Paola Pinilla}
\affiliation{Mullard Space Science Laboratory, University College London, Holmbury St Mary, Dorking, London, UK}

\author{Leonardo Testi}
\affiliation{Dipartimento di Fisica e Astronomia, Università di Bologna, Via Gobetti 93/2, 40122, Bologna, Italy}

\begin{abstract}
Free-floating planetary-mass objects (FFPMOs) are known to harbor disks at young ages. Here, we present 1-13$\,\mu m$ spectra for eight young FFPMOs with masses of 5-10 M$_\mathrm{Jup}$ (at ages of 1-5\,Myr), using the NIRSpec and MIRI instruments on the James Webb Space Telescope. We derive fundamental properties of these targets, and find spectral types of M9.5 to L4, with effective temperatures of 1600-1900\,K. The photospheric spectra of our targets show a clear diversity at similar temperatures, especially in the 3-5$\,\mu m$ range, unaccounted for by existing atmospheric models. We find a silicate absorption feature in the photosphere of one of our targets, the first such detection in very young FFPMOs, indicating silicate clouds in their cool atmospheres. Six of our objects show mid-infrared excess emission above the photosphere, as well as silicate emission features, demonstrating the presence of disks. The shape and strength of the latter features constitute strong evidence of grain growth and crystallization, similar to what is seen in more massive brown dwarfs and stars. We also detect emission lines from hydrocarbon molecules in the disks of several targets. These are the lowest mass isolated objects found so far with silicate and hydrocarbon emission features arising in their disks. The presence of disks and their characteristics point to the potential for the formation of rocky companions around free-floating planetary-mass objects.

\end{abstract}


\section{Introduction} 
\label{sec:intro}

In the year 2000, two groups reported discoveries of the first isolated objects with masses below the Deuterium burning limit \citep{zapatero2000,lucas2000}. They were found as members of the $\sigma$\,Orionis and the Orion Nebula young clusters, as a low-mass extension of the population of free-floating brown dwarfs. Not being able to ignite any fusion reactions, their masses are in the same realm as those of giant planets, with corresponding spectral types of very late M or early L (at young ages). Since then, samples of such free-floating planetary-mass objects (FFPMOs) have been found in several star-forming regions surveyed with sufficient depth (\citealt{{luhman2007,scholz2012,miretroig2022,martin2024}}). Similar objects have also been found in young stellar associations and in the field (\citealt{delorme2012,schneider2016}).

The existence of FFPMOs raises several important questions about the processes that form stars and planets. For example: (1) What is the lowest mass limit for an object to form like a star?; (2) What physical conditions foster or inhibit the formation of such objects?; (3) How many giant planets are ejected early in their evolution to form a population of rogue planets?; and, finally, (4) can FFPMOs form their own miniature planetary systems (and should these companions be referred to as planets or moons)? Observationally, addressing these questions requires both the identification of FFPMOs in diverse regions and detailed characterization of their properties. These goals can be accomplished with the aid of the supreme infrared sensitivity of the James Webb Space Telescope (JWST) and its suite of excellent instruments \citep{rigby2023}.

For more than two decades, it has been established that disks are found around brown dwarfs with masses around and below the Deuterium burning limit (\citealt{natta2001, jaya2003a, luhman2005,scholz2008,damian2023b}). The primary evidence for the presence of disks around FFPMOs came from the excess flux observed in the infrared 3-8$\,\mu m$ bands compared to the photosphere. Further characterisation of these FFPMO disks was hampered by the low luminosity of such targets. Prior to JWST, only one FFPMO had a (low signal-to-noise ratio) mid-infrared spectrum \citep{joergens2013}; the same object also has a sub-mm detection \citep{bayo2017}. With the enhanced mid-infrared sensitivity of JWST, we are now in a position to improve upon these earlier detections and to conduct a more comprehensive survey of the FFPMO disks. 

We have observed eight planetary-mass objects suspected of hosting disks with JWST, obtaining spectra from 1 to 13$\,\mu m$. For one of these objects, CHA1107-7626, we have recently published JWST spectra, highlighting the presence of hydrocarbon molecular emission lines in the mid-infrared \citep{flagg2025}. In this paper, we present a full analysis of JWST infrared spectra for the entire sample, focused on the empirical characterisation. In Section \ref{sec:data}, we present the targets, the dataset, and the reduction. Section \ref{sec:nirspec} focuses on the analysis of the near-infrared data from 1 to 5$\,\mu m$. The mid-infrared spectra from 5 to 12$\,\mu m$ are discussed in Section \ref{sec:miri}. We discuss and summarise the outcomes of our work in Section \ref{sec:Summary}.

\section{The JWST dataset}
\label{sec:data}

\subsection{Target selection}

The goal of this project was to investigate potential disks around isolated objects with masses significantly below the Deuterium burning limit. For this purpose, we searched the literature for objects that are a) members of young (1-3\,Myr), nearby ($<200$\,pc) star-forming regions, b) have spectral types of early L, indicating masses of 5-10$\,M_{\mathrm{Jup}}$ for such young ages, and c) show signs of infrared excess emission in \textit{Spitzer}/IRAC (Infrared Array Camera) photometry. We identified eight good candidates, in three star-forming regions - four in Taurus, three in Chamaeleon I, and one in $\rho$ Ophiuchus - listed in Table \ref{tab:pmo_prop}. All eight are among the lowest mass isolated objects with some evidence for infrared excess, all identified as members of star-forming regions. For all eight, a ground-based near-infrared spectrum and photometry are available in the literature, see Table \ref{tab:pmo_prop} for references. In this paper we provide a new determination of their fundamental properties using the higher-quality data obtained from JWST.

The distances to our FFPMOs was estimated using \textit{Gaia} DR3 (\citealt{gaiamission2016,gaia2023}). Our targets are themselves too faint to be detected by \textit{Gaia}; however, many stellar members of the same star-forming regions have well-determined parallaxes. For each target, we identified around 8-10 young stars in the same star-forming region which are closest to the target, using cluster membership catalogues by \cite{esplin2017} for Chamaeleon I, \cite{esplin2019} for Taurus, and \cite{esplin2020} for $\rho$ Ophiuchus. We adopted the mean distance and the standard deviation of these stellar neighbours derived from the \textit{Gaia} DR3 parallax values as proxy for the distance and corresponding uncertainty to the given target. These distances are presented in Table \ref{tab:pmo_prop}. The estimated distances from \textit{Gaia} DR3 are consistent within the uncertainty with the distances measured by \cite{bailerjones2021}.

\begin{table*}
\centering
\caption{The list of targets observed for this paper, in order of right ascension, with newly determined properties. The prefix 'UGC' indicates targets belonging to the Taurus region, 'CHA' for targets in Chamaeleon I and 'UHW' for the target in $\rho$ Ophiuchus. The spectral type (SpT), effective temperature (T$_\mathrm{eff}$), extinction (A$_V$) and radius are derived in Section~\ref{sec:nirspec}. The distance to each target is estimated in Section~\ref{sec:data}.} 
\label{tab:pmo_prop}
\begin{tabular}{lcccccccccccl}
\noalign{\smallskip}
\hline
\noalign{\smallskip}
Object & RA    & Dec   & SpT & SpT & T$_\mathrm{eff}$ & A$_V$ & A$_V$ & Distance & $M_J$ & Radius  \\
 & & & Lit & This work & & Lit & This work & & & \\
       & (deg) & (deg) &  &   & (K)  & (mag) & (mag) &(pc)     & (mag) & (R$_\mathrm{Jup}$)\\
\noalign{\smallskip}
\hline
\noalign{\smallskip}  
UGC0417+2832$^a$     & 64.4916 & +28.5426 & M9-L7 (L3) & L4 & 1600 & 3.67 & 4.3 & 126$\pm$5 & 12.91 &1.8\\
UGC0422+2655$^a$     & 65.5057 & +26.9201 & L1 & L2 & 1700 & 2.07 & 2.4 & 163$\pm$3 & 11.39 &2.9 \\
UGC0433+2251$^a$     & 68.4753 & +22.8553 & L0.5 & L2 & 1600 & 1.56 & 1.3 & 157$\pm$9 & 11.63 &2.9 \\
UGC0439+2642$^a$     & 69.7823 & +26.7100 & M9.5-L4 (L1) & M9.5  & 1800 & 2.07 & 5.6 & 139$\pm$6 & 11.00 &2.5\\
CHA1107-7626$^b$     & 166.7820 & -76.4424 & M9-L1 (L0) & M9.5 & 1800 & 0.00 & 1.4 & 195$\pm$4 & 10.79 &2.8 \\ 
CHA1110-7633$^c$     & 167.6743 & -76.5518 & M9-L3 (L0) & M9.5 & 1800 & 7.52 & 7.0 & 193$\pm$2 & 10.12 &3.8 \\ 
CHA1110-7721$^c$     & 167.7085 & -77.3649 & M9-L2 (L0) & L2   & 1700 & 5.37 & 4.6 & 190$\pm$6 & 11.09 &3.1 \\ 
UHWJ247.95-24.78$^d$ & 247.9543 & -24.7806 & L1 & L1 & 1900 & 0.00 & 2.6 & 145$\pm$5 & 11.15 &2.2 \\ 
\noalign{\smallskip}
\hline 
\noalign{\smallskip}
\end{tabular}
\begin{flushleft}
References for spectral type and extinction: (a) \cite{esplin2019}, (b) \cite{luhman2008}, (c) \cite{esplin2017}, (d) \cite{allers2020}. The spectral types given within the brackets indicate the adopted values in the respective works. The A$_J$ values from literature were transformed to A$_V$ using the \citet{gordon2023} relation.
\end{flushleft}
\end{table*}

\begin{figure}
    \centering
    \includegraphics[width=\linewidth]{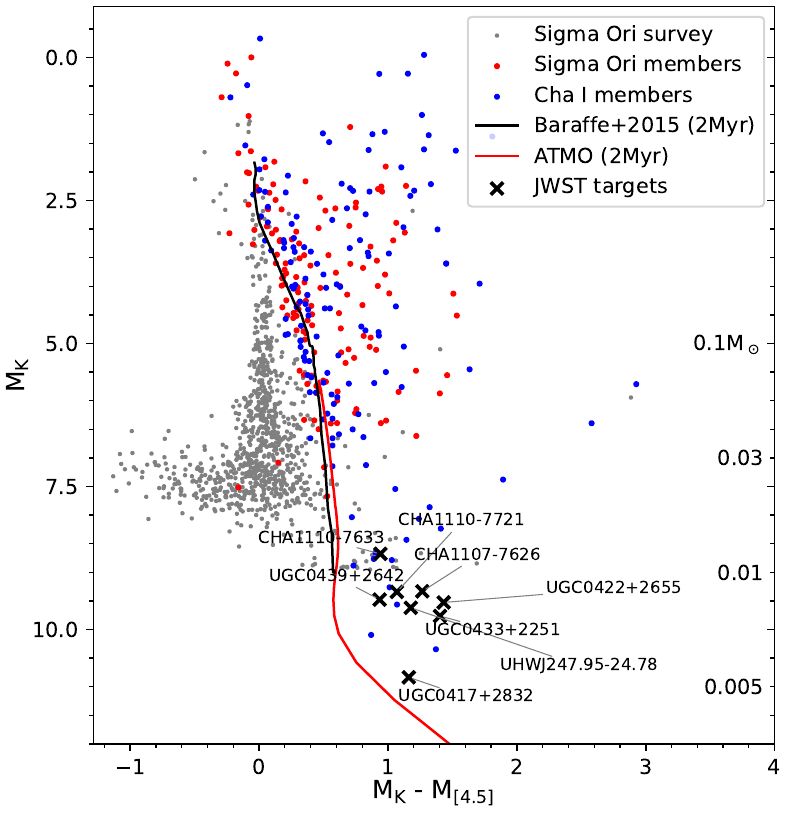}
    \caption{A $K-4.5\,\mu m$ color-magnitude diagram highlighting the location of our objects observed with JWST along with the distribution of low-mass members from similar age $\sigma$ Orionis \citep{damian2023a} and Chamaeleon I \citep{esplin2017} star-forming regions. The magnitudes are in absolute scale, transformed using their respective extinction and distance to the cluster (see Table \ref{tab:pmo_prop} for values). The black and red curves show the 2 Myr isochrone from \cite{baraffe2015} and \citet{phillips2020}, respectively. Masses corresponding to the K-band magnitudes are on the right.}
    \label{fig:cmd}
\end{figure}

In Figure \ref{fig:cmd} we compare the infrared colours of our targets with those of young stars, brown dwarfs, and planetary-mass objects. The figure shows an infrared colour-magnitude diagram for two star-forming regions, compiled from the literature, with our eight targets overplotted and labelled. The $K$ and 4.5$\,\mu m$ magnitudes for our targets are from \citet{luhman2008}, \citet{esplin2017}, \citet{esplin2019} and \citet{esplin2020}. The magnitudes were transformed to absolute magnitudes using the $A_V$ and distance from Table~\ref{tab:pmo_prop} and the \citet{gordon2023} (refer \citealt{fitzpatrick1999,gordon2009,gordon2021,decleir2022} used by \citealt{gordon2023}) extinction relations. The figure also includes isochrones from two models for an age of 2\,Myr from \citet{baraffe2015} and \citet{phillips2020}. The plot illustrates that according to their absolute magnitude these objects have masses below or around $0.01\,M_{\odot}$. In addition, their $K-4.5\,\mu m$ colour is enhanced compared to the photospheric colour, likely due to the presence of a surrounding disk. 

\subsection{Observations and reduction}
All eight targets were observed as part of program 4583 in JWST cycle 3 between August 21 and October 8 2024. We used the prism mode on the Near-Infrared Spectrograph (NIRSpec) and the low-resolution spectroscopy (LRS) mode on the Mid-Infrared Instrument (MIRI) to cover the spectrum from 1 to 13$\,\mu m$ at a low resolution of $\sim 100$. The setup was identical for all 8 targets.

For NIRSpec, we used the S200A1 slit, the full subarray, and the NRSIRS2RAPID readout mode. With 3 dithers, 3 integrations per exposure, and 4 groups per integration, the total on-source time was 656.5\,s, for each target. For MIRI, we used the full subarray, with dithers along the slit, and FASTR1 readout pattern. The total on-source time per target was 3591\,s, split into 2 dithers, 8 integrations per exposure, and 80 groups per integration. When designing the program, the goal was to achieve a signal-to-noise ratio above 100 for the 1-3$\,\mu m$ domain, and above 10 up to $13\,\mu m$. In total, the program used 22.2\,h of telescope time.

\begin{figure*}
    \centering
    \includegraphics[width=\textwidth]{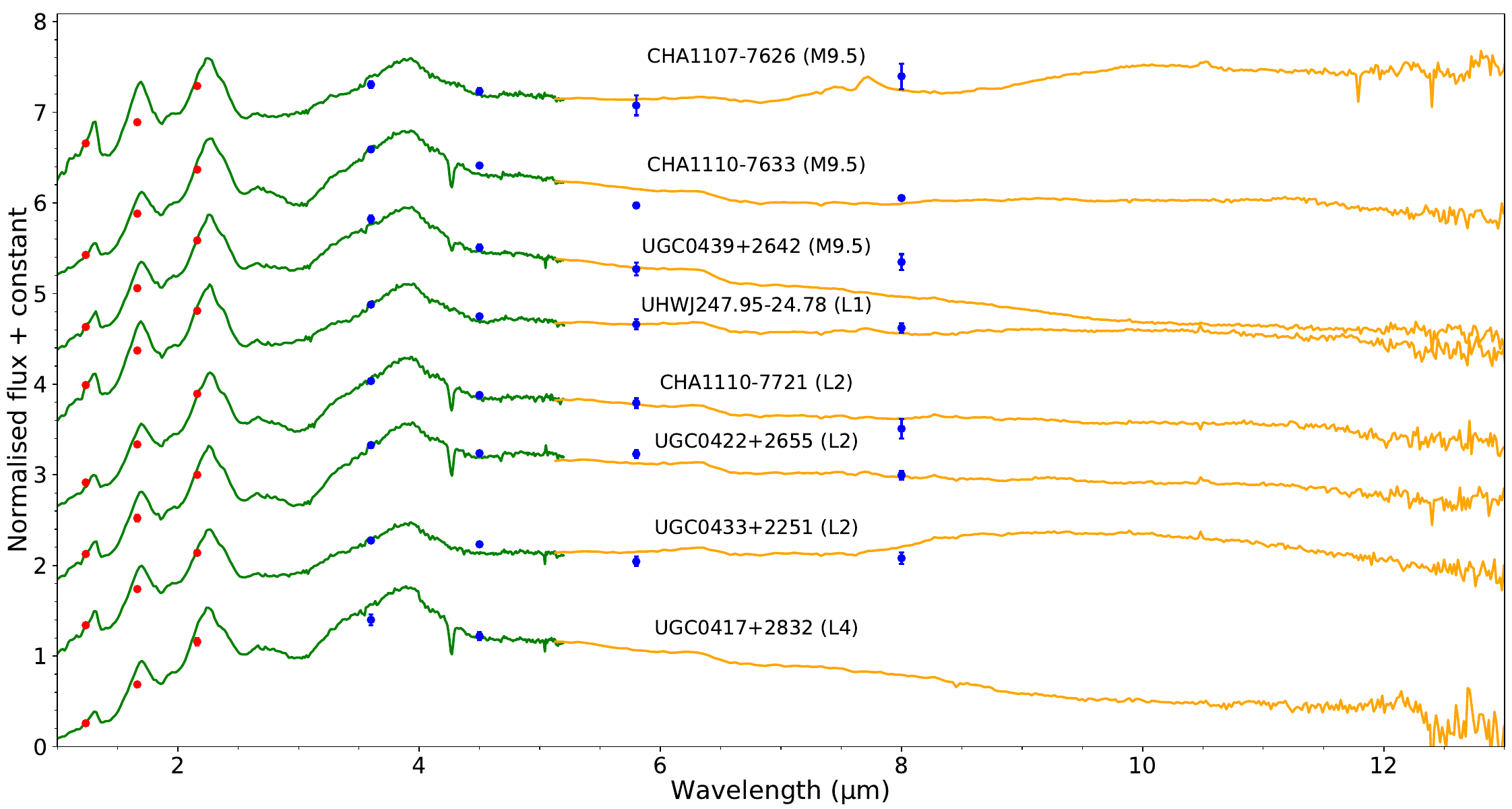}
    \caption{Observed spectra of all the eight objects with JWST NIRSpec (green) and MIRI (orange) normalised to the flux at 2.5$\mu$m. Published photometry at near-infrared (red) and mid-infrared wavelengths (blue) is overplotted.}
    \label{fig:8spectra}
\end{figure*}

In the following, we use the JWST spectra \footnote{The JWST data presented in this article were obtained from the Mikulski Archive for Space Telescopes (MAST) at the Space Telescope Science Institute. The specific observations analyzed can be accessed via \dataset[doi:10.17909/xqx9-v187]{http://dx.doi.org/10.17909/xqx9-v187}} produced with pipeline version 1.15.1 \citep{bushouse2024} with the default settings. The CRDS versions and context used by the JWST pipeline were 12.0.9 and jwst\_1322.pmap, respectively. We found that the pipeline products are of excellent quality. Apart from cosmetic removal of a couple of outlying data points, we did not introduce any further reduction steps. From the MIRI 5$\,\mu m$ verification images, we see that all our targets appear to be isolated single source objects. In Figure \ref{fig:8spectra} we show the complete reduced data set for our eight targets, including the NIRSpec (green) and MIRI (orange) spectra, along with the available infrared photometry. The NIRSpec and MIRI spectra match each other well at 5$\,\mu m$. In general, the photometry also matches the spectra, especially considering that the photometry is over 10 years old and young objects are expected to show some level of variability. The spectra show minimal noise for most of the covered wavelength range. We note that beyond 11$\,\mu m$ the noise level increases, but we can still make useful flux estimates up to 13$\,\mu m$. 

\section{The photospheric spectra}
\label{sec:nirspec}

In this section we analyze the NIRSpec spectra from 1-5$\,\mu m$, focused on deriving the photospheric properties of our targets. 

\subsection{Spectral type fitting}

We first determine the spectral type and extinction by fitting our spectra (from 1 to 2.5$\mu$m) to a grid of reddenened standard templates. Our methodology is similar to that used by \cite{damian2023a} and \cite{langeveld2024}. The set of standard spectral templates we have used are as follows,
\begin{enumerate}
    \item M0-L0 (at intervals of 0.5 subclass), L2, L4 and L7 templates from \cite{luhman2017}. 
    \item L1 and L6 from \cite{allers2013} for objects 2MASS J05184616–2756457 and 2MASS J22443167+2043433 respectively.
    \item L3 from \cite{allers2013} constructed as an average of three objects proposed in \cite{cruz2018}.
    \item L5 proposed by \cite{piscarreta2024} for 2MASS J21543454-1055308, with the spectrum obtained from \cite{gagne2015}. 
    \item M4-L9 (for every 1 subclass) old field dwarfs from the SpeX prism spectral library \citep{burgasser2014}.
\end{enumerate}
To indicate the age of the templates, we use the prefix 'Y' for young and 'F' for old field dwarfs. All the templates listed above with numbers 1, 2, 3, and 4 are considered as young dwarfs ('Y'): M0-L0, L2, L4, L7 have ages $<$5 Myr; L1, L3, L5 and L6 have ages of $\sim$10-100 Myr (refer \citealt{allers2013} and \citealt{piscarreta2024} for details) and those under number 5 are considered as old dwarfs ('F').

\begin{figure*}
    \centering
    \includegraphics[width=0.75\textwidth]{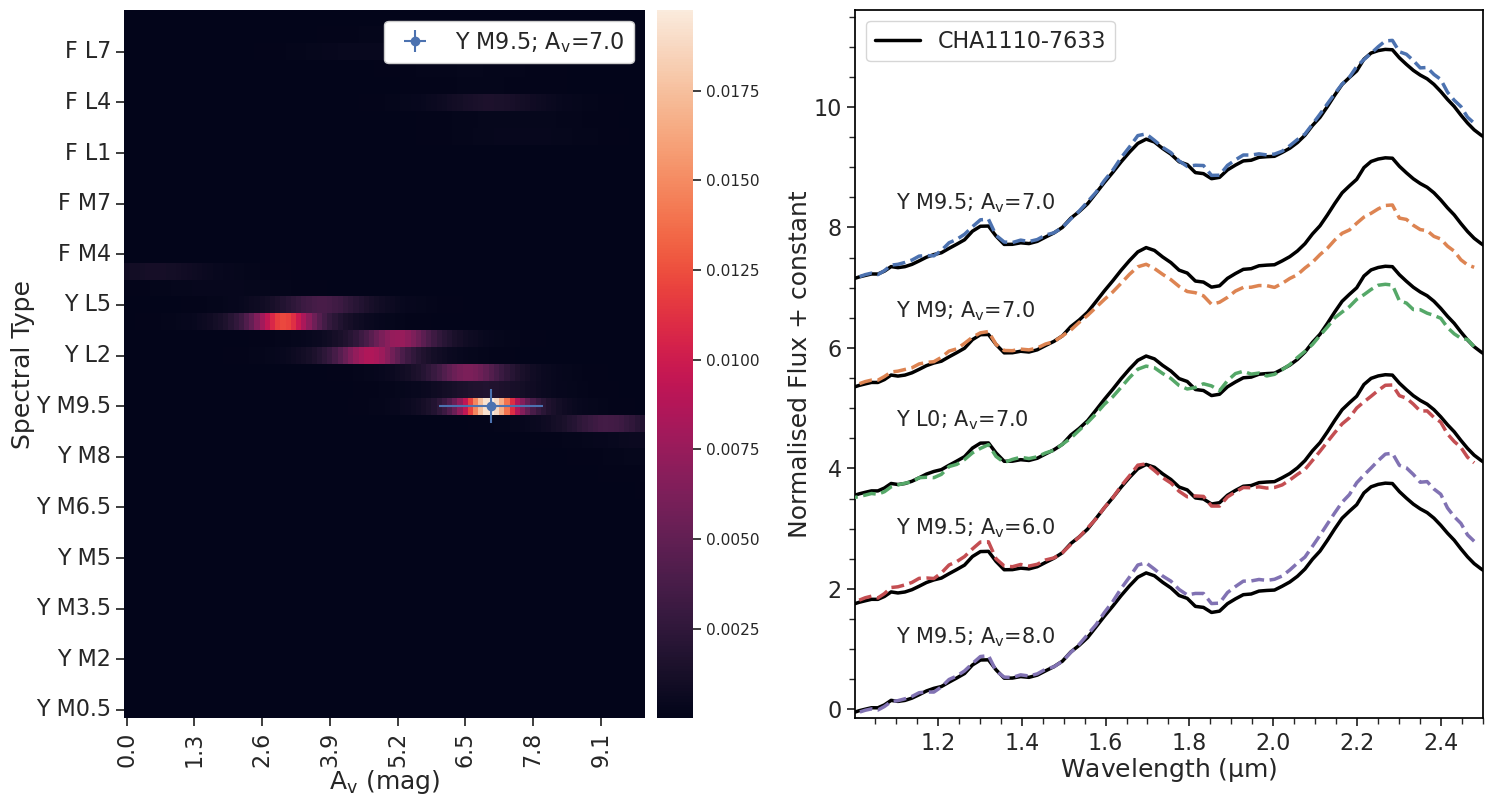}
    \caption{(left) A$_V$ vs spectral type map of one of our targets (CHA1110-7633). The order of templates in the y-axis has no significance and are only grouped according to their ages. The spectral types prefixed with 'Y' and 'F' indicate the young dwarfs and old field dwarfs, respectively. The colorbar indicates the normalised 1/($\chi^2$)$^2$ value where the lowest $\chi^2$ corresponds to the brightest color. The blue colored marker denotes the best-fitting spectral type and A$_V$ with an uncertainty of 0.5 spectral subtype and 1 mag in extinction (refer text for details). (right) NIRSpec spectra of the same object normalised to the flux at 1.5 $\mu$m overplotted with the best fit template and extinction (blue) re-sampled to match the target resolution. The templates with $\pm$0.5 subtype and $\pm$1 mag in extinction are shown in different colors for comparison.} 
    \label{fig:spectral_typing}
\end{figure*}

We calculate the $\chi^2$ goodness-of-fit of our object spectra with the grid of templates reddened by an A$_V$ between 0 to 10 mag in steps of 0.1 mag using the \cite{fitzpatrick1999} extinction law. The choice of extinction law does not have any significant effect on the result, as reported in \cite{almendros2022}. In Figure \ref{fig:spectral_typing} we show the results for one example -- the fitting produces a very good match for a relatively narrow spectral type and extinction range.  The best fitting spectral type and $A_V$ are reported in Table \ref{tab:pmo_prop}. We visually examine the parameter range close to the best fit templates, by comparing with our spectra. In Figure \ref{fig:spectral_typing} we show as an example the spectrum of CHA1110-7633 compared with the best fit, and also shifted in subtype by $\pm 0.5$ and in $A_V$ by $\pm 1$\,mag. From this comparison, we infer typical uncertainties of 0.5 subtype and 1\,mag in $A_V$. 

We find that all our targets fit well with young templates for spectral types of M9.5 to L4, as expected for planetary-mass members in star-forming regions. For four of our targets our estimates are 1-2 spectral types later than previously reported in the literature (see Table~\ref{tab:pmo_prop}) and for three targets its 0.5-1.5 subtypes earlier.  We observe that all our targets have a prominent triangular-shaped peak in the H-band that is characteristic of young late M-type and cooler objects \citep[e.g.,][]{jaya2006, scholz2012}. The extinction $A_V$ ranges from 1.3 to 7\,mag. The $A_V$ values estimated above are consistent for most targets with values previously reported in the literature with differences $<$0.5-1.5 mag (see Table~\ref{tab:pmo_prop}). The two exceptions are UGC0439+2642 and UHWJ247.95-24.78 for which our values are $\sim$2.5-3.5 mag higher than previous estimates by \citet{esplin2019} and \citet{allers2020}, respectively.

\subsection{Comparison with model spectra}

\begin{figure}[h]
    \centering
    \includegraphics[width=\linewidth]{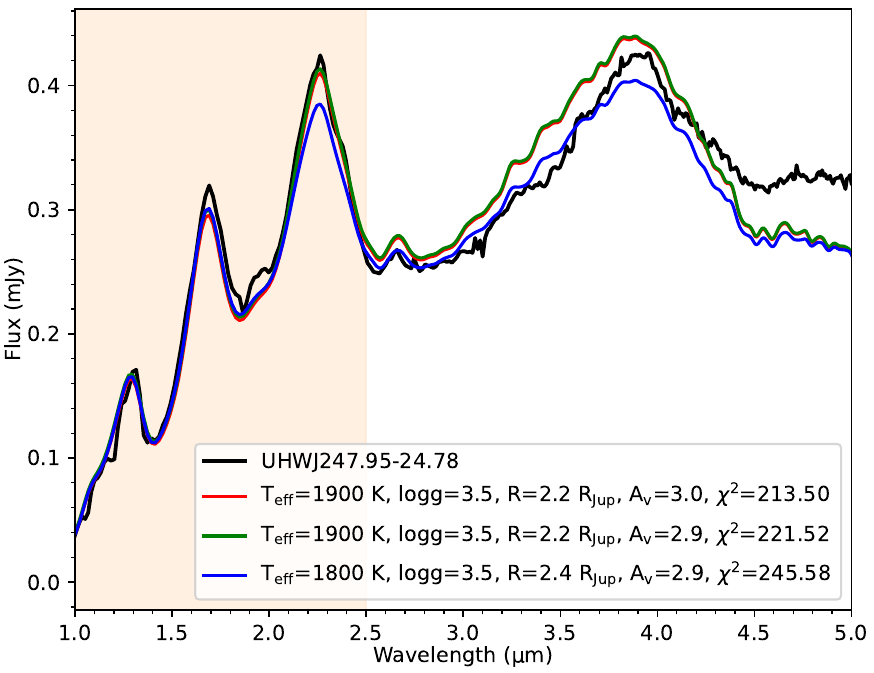}
    \caption{NIRSpec spectrum of UHWJ247.95-24.78 overplotted with the three best fitting photospheric models reddened and scaled by the corresponding A$_\mathrm{V}$ and radius as indicated. The wavelength range considered for the comparison is highlighted by the shaded region. We find that the object fits best with the model spectra for an effective temperature of 1900 K and A$_\mathrm{V}$ of 3 mag.}
    \label{fig:model_fitting}
\end{figure}

We fit the NIRSpec data from 1 to 2.5$\,\mu m$ with the BT-Settl photospheric models \citep{allard2012}. We exclude the 2.5-5$\,\mu m$ range to avoid possible contamination from excess flux due to disk emission. The free parameters in this process are the effective temperature, $\log{g}$, extinction, and the radius in the dilution factor ($R^2/D^2$; where $R$ is the radius and $D$ is the distance). We vary the radius between 0.5 to 4 R$_\mathrm{Jup}$ and the $A_V$ between 0.5 mag above and below the value derived from the spectral type fitting. We assume the distances as listed in Table \ref{tab:pmo_prop}. The reduced $\chi^2$ is estimated using the following relation,
\begin{equation}
   \chi^2 = \frac{1}{N-4} \sum_{i=1}^{N} \frac{(O_i - E_i)^2}{{\sigma_i}^2}
   \label{eq:reduced_chi_squared}
\end{equation}

\noindent where, \textit{N} is the number of data points, \textit{O} is the observed target flux, \textit{E} is the model flux, and \textit{$\sigma$} is the uncertainty in the observed flux. We use the recent extinction relation from \citet{gordon2023} to redden the model spectra. In Figure \ref{fig:model_fitting} we present the results for one of our targets (UHWJ247.95-24.78).

\begin{figure*}
    \centering
    \includegraphics[width=\linewidth]{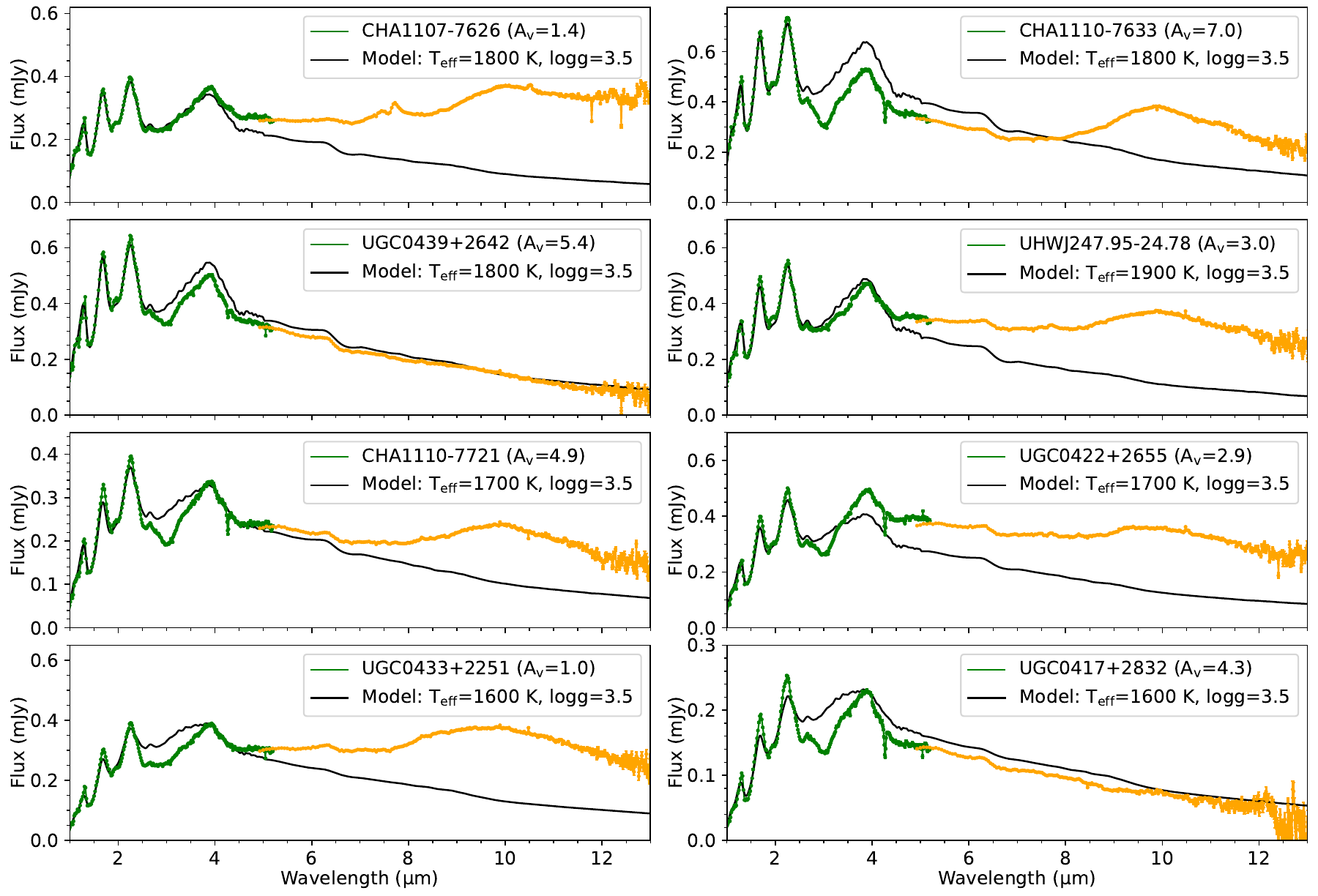}
    \caption{Complete NIRSpec and MIRI spectra of all eight targets dereddened by the extinction estimated through the model fitting along with the best fit photospheric model from BT-Settl \citep{allard2012} scaled by the corresponding dilution factor (see text for details). }
    \label{fig:all_spectra_with_models}
\end{figure*}

For most objects, the BT-Settl models give an acceptable fit to the JHK bands. Again, the resulting best-fit effective temperatures and radii are listed in Table \ref{tab:pmo_prop}. The complete spectra (1-13 $\, \mu m$) of all our targets along with their respective photospheric best fit models are shown in Figure \ref{fig:all_spectra_with_models}. Given that the model spectra are only available in steps of 100\,K, the typical uncertainty in T$_\mathrm{eff}$ will be at least 100\,K. We find all our targets to match best with a $log{\ g}$ of 3.5, consistent with the expected value for young low-mass objects. We note that some of our T$_\mathrm{eff}$ estimates tend to be slightly lower compared to published conversions between spectral type and temperature. For brown dwarfs in young moving groups with spectral types M9.5 to L4, we would expect temperatures of 1600 to 2100\,K \citep{sanghi2023}. These objects would be slightly older than our targets. For very young star-forming regions, the conversion between spectral type and temperature is poorly constrained; at late M spectral type we would expect temperatures below 2500\,K \citep{muzic2014}.

We emphasize that the derived temperatures are model-dependent (as effective temperatures always are), and should only be used in conjunction with the given model. \citet{sanghi2023} and \citet{hurt2024} observed that the BT-Settl atmospheric models produce lower temperatures in some dwarfs in the M-L transition region when compared to temperatures derived from evolutionary models. This results in a pile-up of objects at $\sim$1800 K, creating gaps in the spectral type vs temperature relation at around $\sim$2000 K.  Therefore, using a different wavelength regime or a different model family to compare the spectra would likely yield minor differences in the results. 

\begin{figure}[h]
    \centering
    \includegraphics[width=\linewidth]{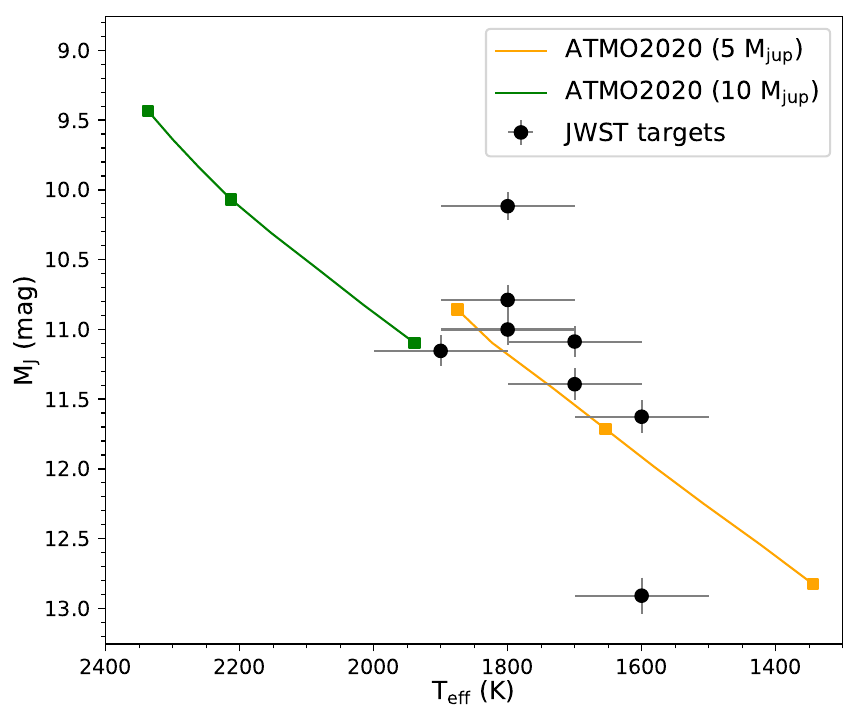}
    \caption{HR diagram for our sample, in comparison with theoretical evolutionary tracks. The figure shows absolute J-band magnitude vs. effective temperature, both listed in Table \ref{tab:pmo_prop}. We overplot error bars assuming errors of $\pm 100$\,K in temperature and $\pm 10$\% in brightness. The evolutionary tracks are from ATMO2020 models \citep{phillips2020}, for 0.005 and 0.01$\,M_{\odot}$. The squares on the tracks mark ages of 1, 2, 5\,Myr, from top to bottom.}
    \label{fig:hrd}
\end{figure}

In Figure \ref{fig:hrd} we show a Hertzsprung–Russell (HR) diagram comparing our objects with evolutionary tracks for 0.005 and 0.01$\,M_{\odot}$ for ages of 1-5\,Myr. We use the ATMO 2020 CEQ models for the 2MASS filter system from \citet{phillips2020}. For this purpose we calculate absolute J-band magnitude from the available photometry, using the distance as derived above and the extinction derived from spectral typing, converted to J-band using the extinction law from \citet{gordon2023}. For six out of the eight targets, the position in the HR diagram is consistent with the evolutionary track for 0.005$\,M_{\odot}$, for ages of 1-2\,Myr. The exceptions are UGC0417+2832 which is underluminous and CHA1110-7633 which is overluminous compared to this model, but still too cool for the model at higher mass. Notably, CHA1110-7633 is the source with the highest $A_V$ in the sample. Based on the HR diagram, all our objects should have masses conservatively below 0.01$\,M_{\odot}$, for the plausible age range adopted here. We note that this conclusion is robust against minor changes in the model-dependent effective temperatures.

\subsection{Discussion: Photospheric spectra of young planetary-mass objects}

\begin{figure}[h]
    \centering
  \includegraphics[width=\linewidth]{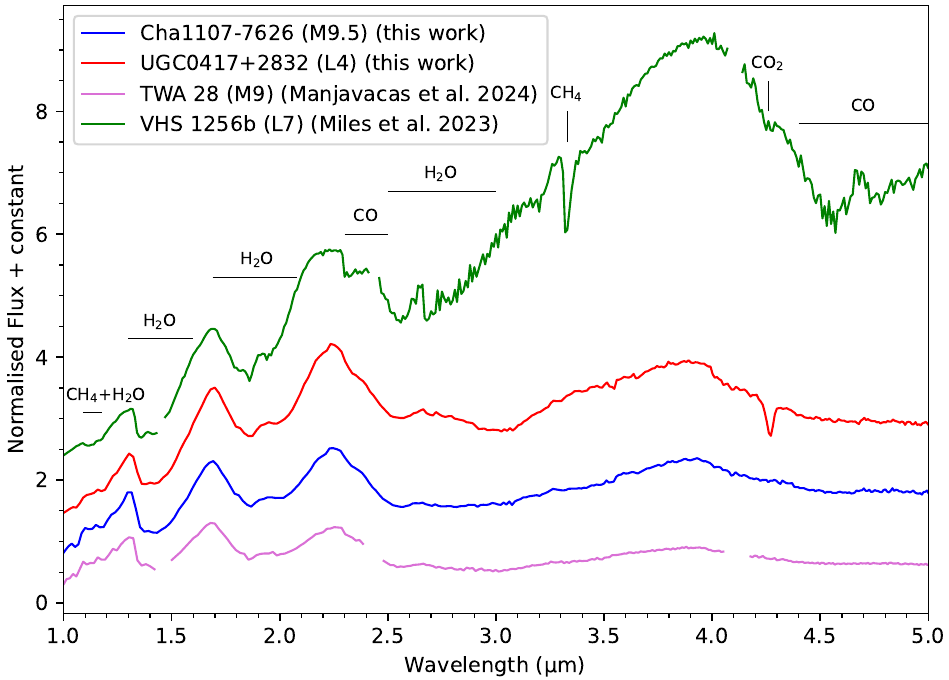}
    \caption{Comparison spectra of two of our targets with similar spectral type objects in literature. The prominent molecular absorption features are highlighted. The spectra of the literature objects are resampled to match the resolution of our targets and are normalised to the flux at 1.5$\,\mu m$.}
    \label{fig:comparison_spectra}
\end{figure}

In Figure \ref{fig:comparison_spectra} we show the 1-5 $\,\mu m$ spectra of two of our targets, CHA1107-7626 (M9.5) and UGC0417+2832 (L4), whose spectral types are representative of the range observed in our sample. We compare their spectra with those of two well-studied objects with similar spectral types, TWA 28 and VHS 1256b. TWA 28 (SSSPMJ1102-3431; \citealt{scholz2005}) is a well-studied brown dwarf member of the TW Hydrae association (age $\sim$10 Myr) with an effective temperature of $\sim$ 2600 K (M9) and mass around 20 M$_\mathrm{Jup}$ (\citealt{venuti2019,manjavacas2024}). \citet{manjavacas2024} observed this object with JWST NIRSpec in medium-resolution (R $\sim$ 2700). \citet{miles2023} carried out similar medium-resolution observations with the same instrument for VHS 1256b, a planetary-mass companion to an M-dwarf binary system at a distance of $\sim$20 pc \citep{dupuy2020} aged $\sim$140 Myr \citep{dupuy2023} with temperature $\sim$ 1200 K (L7) and mass $<$20 M$_\mathrm{Jup}$ \citep{miles2023}. 

As illustrated in the figure, objects with spectral type from mid M-type and beyond show water absorption features in the near- and mid-infrared regimes due to the presence of water vapor in their atmospheres. The depth of these features changes with decreasing temperature \citep{allers2013}. In Figure \ref{fig:comparison_spectra} we see strong water absorption features between 1.3-1.6 $\,\mu m$, 1.69-2.05 $\,\mu m$ and 2.5-3 $\,\mu m$ in all four spectra. We also see prominent CO absorption around 2.4 $\,\mu m$ and 4.4-5 $\,\mu m$ in VHS 1256b, but it is weaker in the other three objects. The CO$_2$ molecular absorption feature at 4.2$\,\mu m$ has been reported only in a few late L and T dwarfs and exoplanets, and is attributed to high metallicity relative to the Sun (\citealt{tsuji2011,sorahana2012,ahrer2023}). Most of our sources, but not all, show a clear absorption dip at this wavelength, of varying strength, something not seen in the models. Accounting for these features requires a more detailed atmospheric analysis beyond the scope of this paper. 

In the 3-5$\,\mu m$ spectral domain, the photospheric emission is dominated by a peak that is shaped by water and carbon monoxide absorption \citep{manjavacas2024}. As seen in Figure \ref{fig:comparison_spectra}, the strength of this peak is obviously a function of temperature. As shown in Figures \ref{fig:model_fitting} and \ref{fig:all_spectra_with_models}, the 3-5$\,\mu m$ range is often poorly matched by the photospheric models. One object in our sample (UGC0422+2655) shows definitive excess in this band, likely caused by disk emission (see Section \ref{sec:miri}). Several others have flux levels below those predicted by the model that matches the spectrum in the 1-3$\,\mu m$ range. Clearly, the shape of this feature in the 3-5$\,\mu m$ domain is not properly represented by the models.

As can be appreciated in Figures \ref{fig:8spectra} and \ref{fig:all_spectra_with_models}, the slope between 3 and 3.9$\,\mu m$ is also slightly different when comparing objects in our sample. In about half of the sample, the flux in this range appears to rise with a constant slope (see for example CHA1107-7626 or UGC0433+2251). Some other objects show a minor 'bump' roughly at 3.3$\,\mu m$ where the slope flattens (see for example UGC0417+2832 and CHA1110-7721). An example for each type is shown in Figure \ref{fig:comparison_spectra}. The presence or non-presence of this 3.3$\,\mu m$ slope change is not related to a specific spectral type or the presence of disk emission visible at longer wavelengths. It may be related to extinction -- the objects which do not show this tentative bump tend to have low extinction. We note that there is a known methane absorption feature around the same wavelength, visible, for example, in VHS 1256b (Figure \ref{fig:comparison_spectra}). However, our targets are thought to be too warm for methane absorption. There are also known PAH emission in this wavelength domain; thus, imperfect background subtraction could play a role in affecting the slope of the spectra \citep{boersma2023}. Recently \citet{luhman2025} observed the presence of a hydrocarbon absorption feature at 3.4$\,\mu m$ in some very young FFPMOs which otherwise appear to be L dwarfs; we do not see such a feature in any of our sources.

The diversity in the photospheric spectra displayed in our sample point to parameters other than temperature and gravity that influence the spectral appearance in this age and mass range. Apart from variations in metallicity as discussed above, another possibility is inhomogeneity in the cloud distribution \citep[e.g.,][]{radigan2014}. In particular we point to the recent works that show variation in near- and mid-infrared colors and also dust cloud opacity depending on the inclination of the rotational axis relative to the line of sight. The dwarfs with higher inclination angles (equator-on) appear cloudier with redder infrared colors than those with lower inclination angles (pole-on) that appear bluer with lower cloud opacity (\citealt{vos2017,suarez2023}). Clearly, more work is needed to understand the diversity of cool substellar atmospheres.

\section{The mid-infrared spectra}
\label{sec:miri}

In this section we focus on the analysis of the mid-infrared spectra obtained with MIRI/LRS. As discussed in Section~\ref{sec:data}, our targets were selected based on indications of infrared excess in the available \textit{Spitzer} photometry. Our comparison with photospheric spectra show that indeed six out of eight definitively show excess emission beyond 4$\,\mu m$ (Section \ref{sec:nirspec} and Figure \ref{fig:all_spectra_with_models}), as discussed below. For the remaining two sources the mid-infrared spectrum is photospheric; they are further investigated in Section \ref{sec:silicateabs}.

\subsection{Mid-infrared excess from disks}
\label{sec:excess}

Six objects in our sample show clear excess emission in the mid-infrared above the photosphere, best appreciated in Figure \ref{fig:all_spectra_with_models}. The additional emission is comparable to or stronger than the photospheric flux around 10$\,\mu m$. The best explanation for this excess is emission from warm dust in disks. The presence of silicate in emission, typically found in disks around young stars, is also evident in these sources, and will be further analyzed in Section \ref{sec:silicateem}. Thus, our spectra unambiguously demonstrate the presence of circum-sub-stellar material around free-floating planetary-mass objects.

Within this sample there is considerable diversity in the shape of the mid-infrared spectra. Apart from the strength and shape of the silicate feature, the slope of the emission also changes from object to object. In order to measure this, we estimate the extinction corrected flux ratio at 8.0$\,\mu m$/3.6$\,\mu m$. This ratio varies from 0.5 (for Cha1110-7633) to 0.9 (for UGC0433+2251). We note that OTS44, a previously studied planetary-mass object with a disk \citep{joergens2013}, has a slightly larger flux ratio of 0.95 (calculated from photometry).

The slope of the mid-infrared SED reflects the degree of flaring of the disk and is a proxy for the evolutionary stage of the disk \citep{cieza2007}. Previous work distinguishes between full disks, transitional disks, and evolved disks based on this metric \citep{teixeira2012}. This sequence is thought to be linked to the dust evolution -- with ongoing grain growth, the dust decouples from the gas and settles to the disk midplane, leading to lower level of excess emission \citep[e.g.,][]{mohanty2004, scholz2007}.

Our objects show the typical signature of evolved disks, with flat or declining fluxes in the 3-8$\,\mu m$ domain. Beyond 10$\,\mu m$ the flux level also declines in all our objects with disks, with the possible exception of Cha1107-7626. Thus, the slope of the SED provides a first indication of a degree of dust settling and grain growth in disks around planetary-mass objects, as is observed in their more massive counterparts \citep{scholz2007}. 

As a side note: With the relatively weak excess emission at 3-8$\,\mu m$, combined with the low photospheric fluxes, it is challenging to reliably distinguish between objects with and without disks from photometry alone. This is illustrated by the fact that two of the objects in our sample show signs of infrared excess in Spitzer/IRAC photometry, and yet clearly do not harbor disks.

\subsection{Silicate emission feature in the disk}
\label{sec:silicateem}

In this subsection we parameterize and quantify the shape and strength of the 10$\,\mu m$ silicate emission feature in our targets with disks. This feature is a way to assess the evolutionary state of the warm dust in the inner disk, as has been demonstrated with ground-based \citep{mohanty2004} and \textit{Spitzer} data for a number of brown dwarfs \citep{apai2005,scholz2007,pascucci2009}. We note that if these objects hold silicate clouds in their cool atmospheres, unlike their higher temperature and mass siblings, this would alter the strength of the observed silicate emission feature emanating from the disk (see Section \ref{sec:silicateabs}).

Interstellar extinction can be a contributing factor to undermining the strength and shape of the silicate emission feature arising from the disk as shown recently by \citet{arabhavi2025}. Hence, for the following analysis we use the extinction corrected spectra of our targets. We deredden the spectra using the \citet{gordon2023} extinction law and the A$_V$ values derived in  Section~\ref{sec:nirspec} (see Table~\ref{tab:pmo_prop}). The choice of the extinction law stems from its availability in the MIR wavelengths and also because the \citet{gordon2023} extinction curve presents the 10$\,\mu m$ feature due to the silicate dust grains in the ISM. We note that the extinction law does not represent the specific line of sight extinction to the targets and the published extinction curves are the averages. Hence the differences in the extinction laws may alter the shape of the silicate feature.

\begin{figure}
    \centering
    \includegraphics[width=\linewidth]{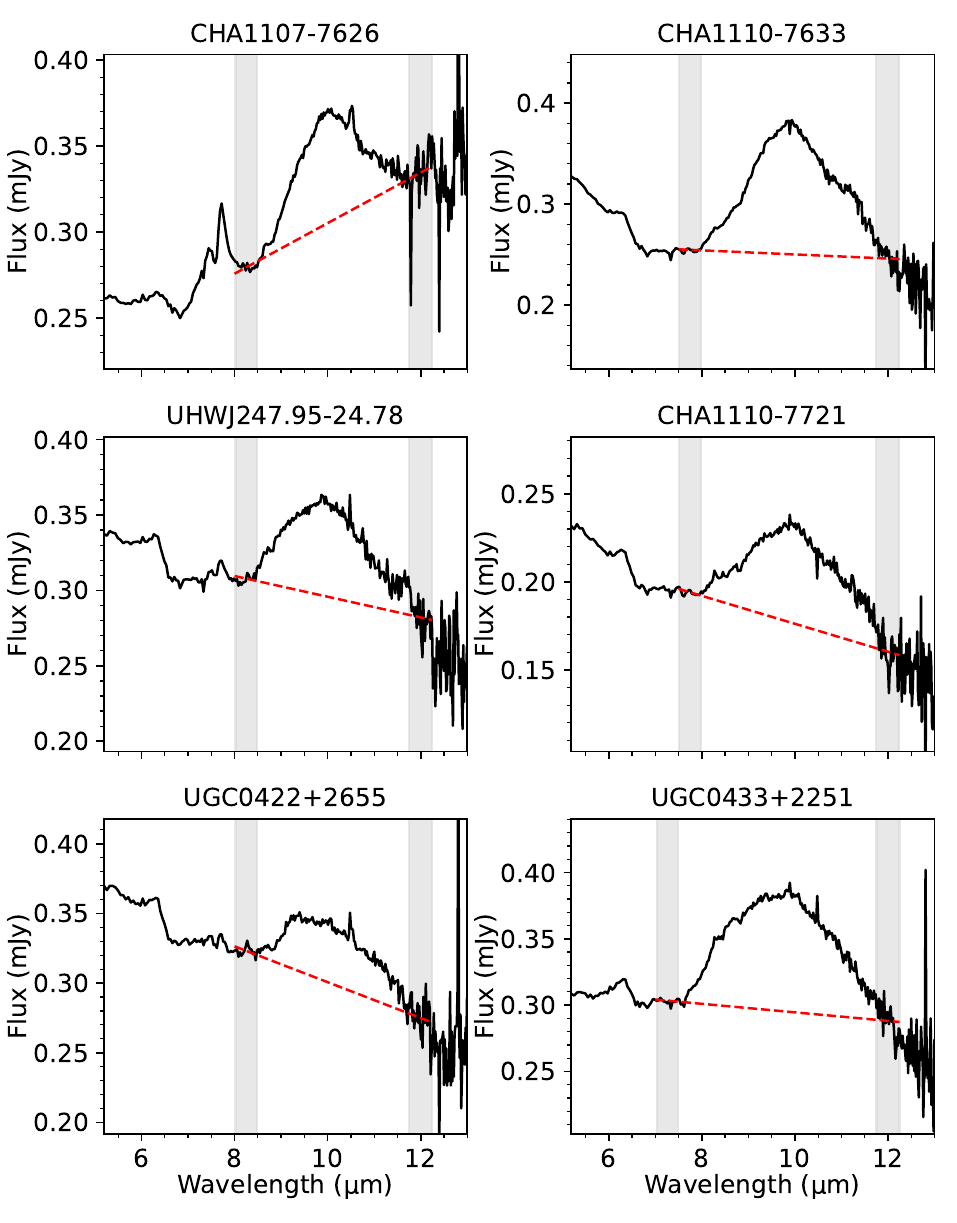}
    \caption{MIRI spectra of our six targets with disk presented individually in each panel dereddened using the A$_V$ values listed in Table~\ref{tab:pmo_prop}. The linear fit to the continuum is shown as a red dashed line and the 0.5$\,\mu m$ wavelength windows used for the fit are shaded in gray. At the shorter wavelength range this was optimized to avoid any emission lines and/or the silicate feature.}
    \label{fig:silicate_feature_continuum}
\end{figure}

To quantify the strength of the silicate feature observed at 10$\,\mu m$, we first trace a linear fit to the continuum around the feature as shown in Figure~\ref{fig:silicate_feature_continuum}. We use a wavelength window of 0.5$\,\mu m$ optimized to each object at the bluer end to exclude any emission lines and/or the shoulder of the silicate feature and around 12$\,\mu m$ at the redder end. The wavelength range used to fit the continuum is highlighted in the figure. All six of our targets with disks show an emission feature above the parameterized continuum. We then normalize the spectra to the continuum and estimate the peak flux over the continuum around 10$\,\mu m$. The prominent emission lines between 9-11.5$\,\mu m$ were masked and the spectra was smoothed to obtain the peak flux above the normalized continuum.

\begin{figure}
    \centering
    \includegraphics[width=\linewidth]{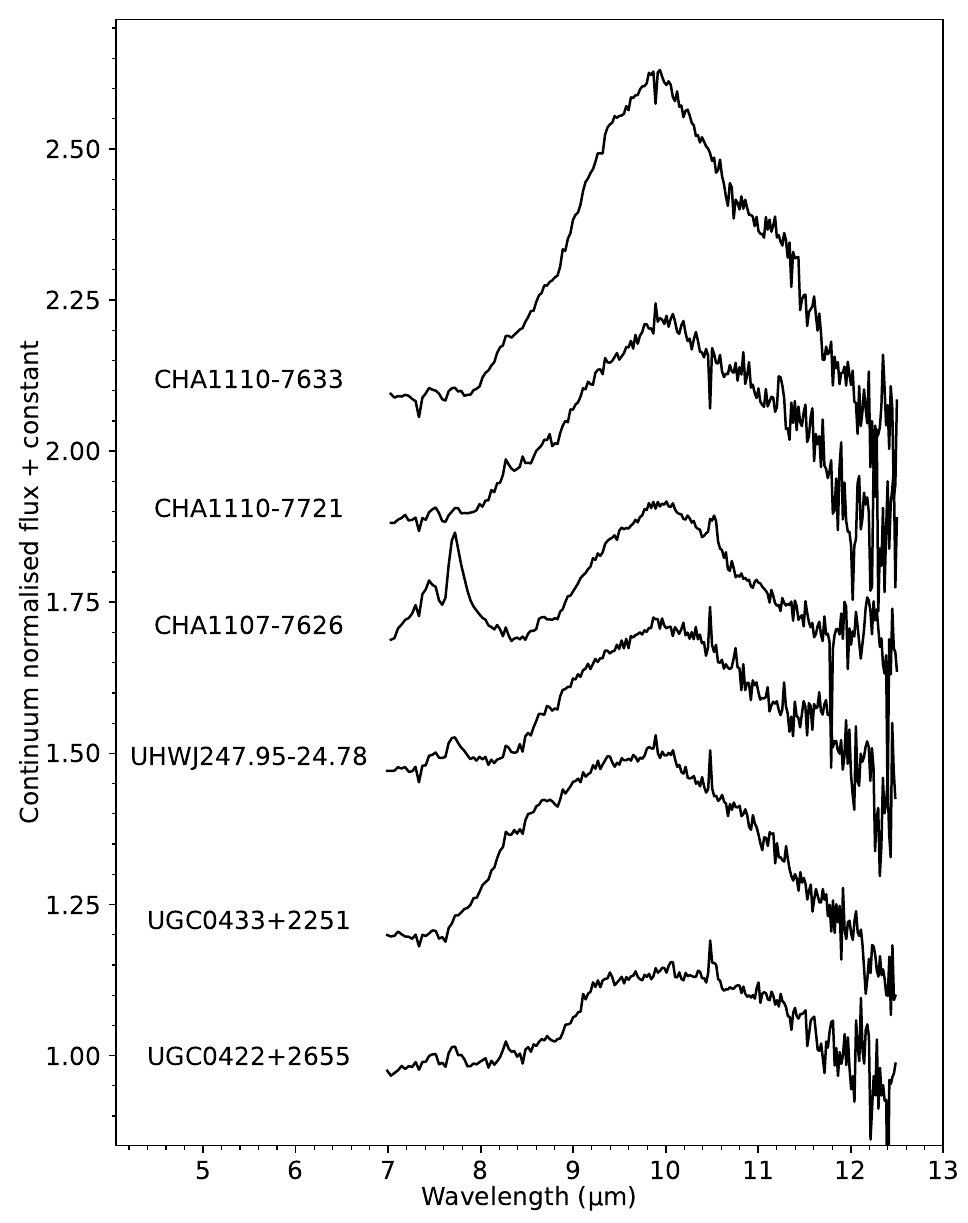}
    \caption{MIRI dereddened spectra normalized to the continuum estimated in the previous figure. We show only the 7 to 12.5$\,\mu m$ region to highlight the silicate emission feature and the diversity in its shape among our 6 objects with disk.}
    \label{fig:silicate_feature_peak}
\end{figure}

The continuum-normalized extinction corrected spectra are presented in Figure~\ref{fig:silicate_feature_peak}, in order of strength of the silicate feature. The plot clearly highlights the diversity in the shape and strength of the silicate feature among our targets, analogous to previous studies for higher-mass brown dwarfs \citep{apai2005}. The variations seen in this figure can be explained by the differences in grain sizes and degree of crystallization. Amorphous ISM-type silicates cause a strong silicate feature, with a single peak around 9-10$\,\mu m$, as seen in the three objects at the top of Figure \ref{fig:silicate_feature_peak}. On the other hand, crystalline-rich silicates typically show two peaks at 9.3 (enstatite) and 11.3$\,\mu m$ (forsterite), resulting in a flattened appearance of the feature, as seen in UGC0422+2655 at the bottom of the plot. For the remaining two objects, the 10$\,\mu m$ feature tends to show a transition state between the amorphous sub-micron sized silicate grain to the processed crystalline silicates. 

\begin{figure}
    \centering
    \includegraphics[width=\linewidth]{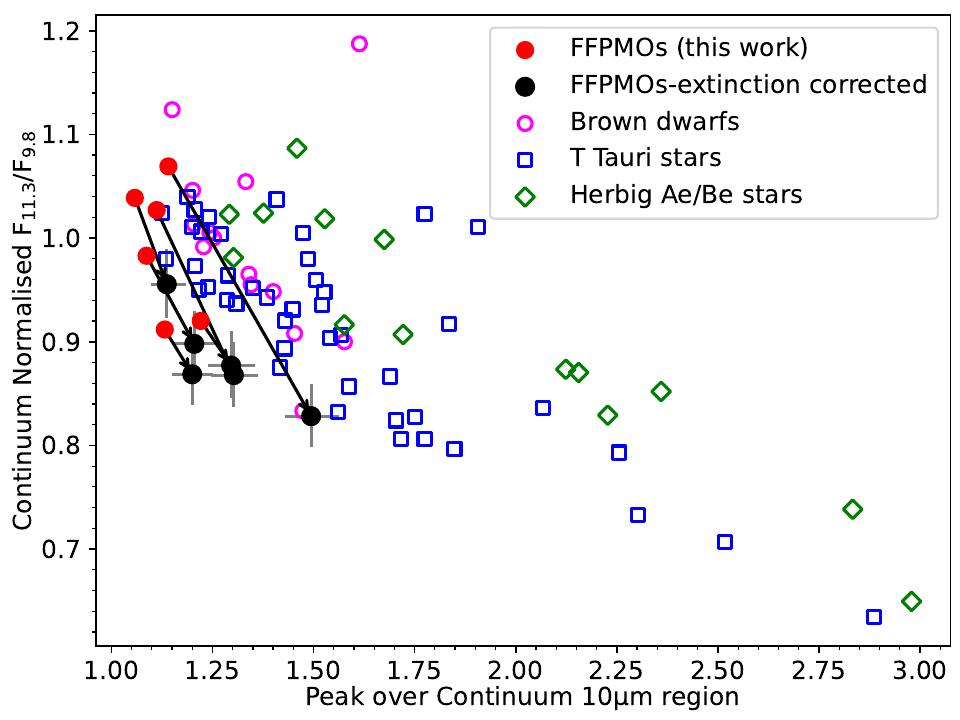}
    \caption{Shape and strength of the silicate emission feature of our FFPMOs with disk (before and after dereddening) and a sample of brown dwarfs, T Tauri stars, and Herbig Ae/Be stars from \cite{pascucci2009}. The error bars indicate the upper and lower values for a change in A$_V$ of $\pm$1 mag. In comparison this shows that the silicates in the disk around FFPMOs are processed to a higher degree than their massive counterparts.}
    \label{fig:silicate_feature_comparison}
\end{figure}

In Figure \ref{fig:silicate_feature_comparison}  we plot the continuum normalized flux ratio at 11.3 vs. 9.8$\,\mu m$ and the peak flux over the continuum for the silicate features. For comparison, we also plot the data for the sample of brown dwarfs, T Tauri stars, and Herbig Ae/Be stars from \citet{pascucci2009}. This comparison sample from \citet{pascucci2009} is likely derived from spectra that are uncorrected for extinction. Hence, we have shown the distribution of our six FFPMOs both before and after extinction correction. The error bars in the figure indicate the uncertainty in the A$_V$ values and do not represent the possible variations due to the extinction laws.

Prior to extinction correction, our six datapoints cover the same parameter range as the previously published brown dwarfs, with low peak over continuum values and flat features (i.e. a flux ratio at 11.3 vs 9.8$\,\mu m$ around 1.0). Dereddening changes primarily the 'colour' of the silicate feature. It shifts the datapoints in the figure downwards. Hence, after extinction correction, our datapoints appear towards the left and below the trend established from the literature.

Overall, the silicate features for some objects in our sample show clear evidence for grain growth and crystallization in disks around planetary-mass objects, similar to what has been observed in higher-mass brown dwarfs. We note that processes that dominate the dust growth such as radial drift velocity are expected to be more efficient around very low-mass objects than brown dwarfs \citep{pinilla2022}. Combined with the evidence for long-lived dusty disks around planetary-mass objects (\citealt{scholz2023,seo2025}), these findings point strongly to the potential for the formation of rocky companions in the disks around free-floating planetary-mass objects.

\subsection{Silicate absorption in the photosphere}
\label{sec:silicateabs}

In field brown dwarfs without disks, a silicate absorption feature at 10$\,\mu m$ has been observed, attributed to the presence of silicate clouds in the cool atmospheres (\citealt{cushing2006,suarez2022} and references therein). Besides brown dwarfs, recently, \citet{miles2023} observed the silicate absorption feature in the MIRI spectra of the planetary-mass companion VHS 1256b. 

\begin{figure*}
    \centering
    \includegraphics[width=0.45\textwidth]{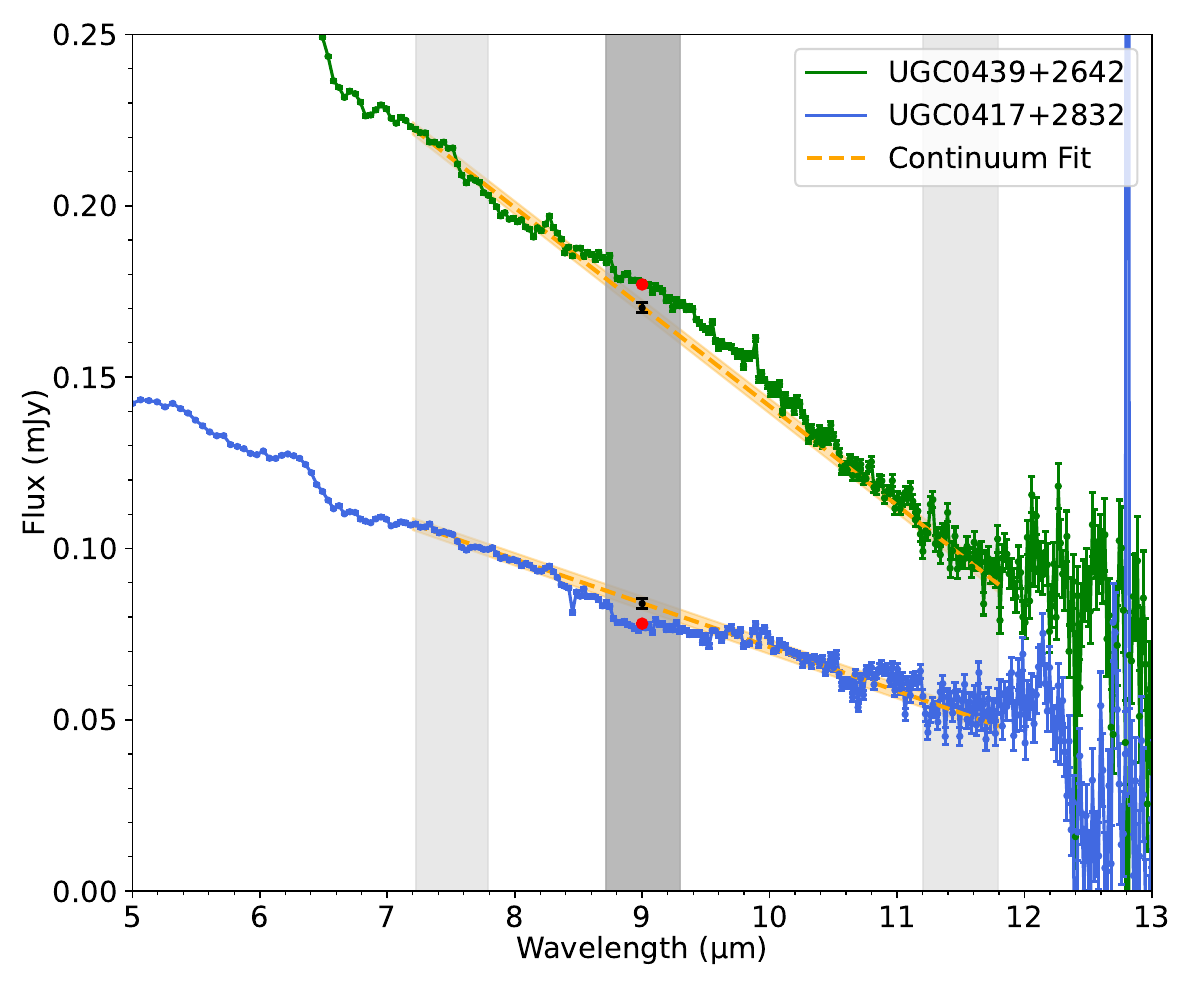}
    \includegraphics[width=0.54\textwidth]{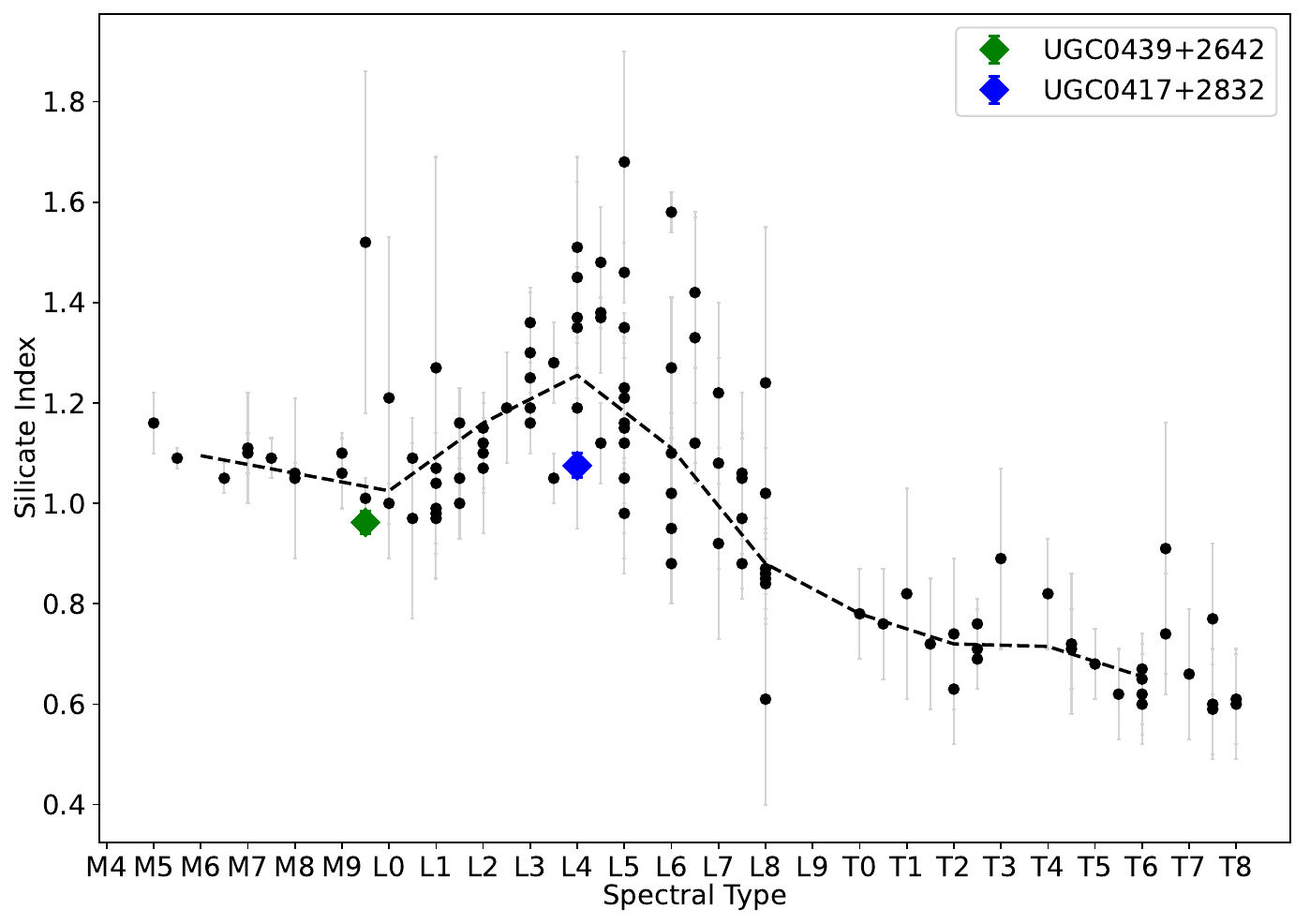}
    \caption{(left) Measurement of silicate index for the two targets in our sample that do not show excess emission above the photosphere in the MIR. Both the spectra are corrected for extinction using the \citet{gordon2023} relation and A$_V$ from Table~\ref{tab:pmo_prop}. The vertical shaded regions of width 0.6$\,\mu m$, highlight the wavelength ranges used to fit the continuum (light gray) and the silicate absorption (dark gray). The orange dashed line and shaded region show the best linear fit to the continuum and the uncertainty, respectively. The black and red points mark the average flux at 9.0$\,\mu m$. (right) Silicate index as a function of spectral type for our two FFPMOs along with the index for dwarfs from \citet{suarez2022}. The error bars for the two FFPMOs indicate the range in silicate index when A$_V$ changes by $\pm$1 mag. The black dashed curve is the median silicate index for bins of two spectral sub-types.}
    \label{fig:silicate_index}
\end{figure*}

For two of our targets, UGC0439+2642 and UGC0417+2832, which lack excess disk emission in the MIRI spectra above the photospheric models (see Figure~\ref{fig:all_spectra_with_models}), we look for evidence of silicate absorption, following a procedure similar to \citet{suarez2022}. We deredden the spectra of these two targets and quantify the silicate index which is the ratio of the continuum flux to the absorption flux at 9.0$\,\mu m$. The continuum at 9.0$\,\mu m$ is defined by the best linear fit to the fluxes in a wavelength window of 0.6$\,\mu m$ centered at 7.5 and 11.5$\,\mu m$ as shown in the left panel of Figure~\ref{fig:silicate_index}. We derive the silicate index by estimating the ratio between the average continuum flux and the average absorption flux in a wavelength window of 0.6$\,\mu m$ centered at 9.0$\,\mu m$. We then compare our results with the silicate index measured for brown dwarfs in \citet{suarez2022} as a function of their spectral types in the right panel of Figure~\ref{fig:silicate_index}. 

We find that one of these two objects, UGC0417+2832 shows silicate in absorption. To our knowledge, this is the first time the silicate absorption has been found in a FFPMO in a young star-forming region (age $<10$ Myr). For the remaining source, UGC0439+2642, the spectrum is slightly above the photosphere. For both objects, the silicate indices are lower than the median index values of the brown dwarfs for a given spectral type. A larger sample will be required to evaluate how the strength of the silicate absorption changes as a function of spectral type at these very young ages.

\subsection{Molecular gas lines}

Most of the sources analyzed here show mid-infrared emission features attributed to molecular gas in their disks. These features are most pronounced in CHA1107-7626. In \citet{flagg2025}, we reported strong methane (7.7$\,\mu m$) and ethylene (10.5$\,\mu m$) emission for this particular object, and discussed similarities in its spectrum with a classical T Tauri star ISO-ChaI 147 observed with JWST/MIRI \citep{arabhavi2024}. The detection of hydrocarbon emission features in a planetary mass object resembling a carbon-rich disk of a low-mass star indicates the similarities in the inner disk conditions across a range of stellar/substellar masses (see also the recent work by \citet{arabhavi2025}).

While the emission line spectrum for CHA1107-7626 is remarkable, it is by no means unique, as can be appreciated from Figure \ref{fig:silicate_feature_peak}. The 10.5$\,\mu m$ ethylene feature is confidently detected in three other disks in our sample: UGC0422+2655, UHWJ247.95-24.78 and UGC0433+2251. We note that there is also a hydrogen 12-8 line at 10.5$\,\mu m$ that occurs in disks. With the low resolution at this wavelength (of about 0.1$\,\mu m$), we are unable to decisively distinguish between hydrogen and ethylene. However, none of our spectra show any evidence of other atomic hydrogen lines \citep{franceschi2024}, which should be present in addition to the 12-8 feature. Therefore we deem it more plausible that the observed 10.5$\,\mu m$ emission is due to ethylene.

Similarly, the 7.7$\,\mu m$ methane feature appears to be be present in CHA1110-7633, UGC0422+2655, UHWJ247.95-24.78, and CHA1110-7721. Other features may be hidden by the low resolution of our MIRI spectra and/or by the excessive noise beyond 12$\,\mu m$. Follow-up studies with higher resolution are required to compile a detailed inventory of the molecular line emission in disks around planetary-mass objects. 

\section{Summary}
\label{sec:Summary}

We present low-resolution 1-13$\,\mu m$ spectra of eight free-floating planetary-mass objects in nearby young star-forming regions observed with the JWST NIRSpec and MIRI instruments. These FFPMOs are members of the young (age $<$5 Myr) regions Taurus, Chamaeleon I, and $\rho$ Ophiuchus with masses of 5-10\,M$_\mathrm{Jup}$. All our targets have been previously observed with ground-based spectroscopy and photometry, with indications of infrared excess emission from disks. With our high-quality JWST spectra, we derive fundamental properties of all eight targets and substantiate the presence of disks around six. For the first time in these young planetary-mass objects, we also report the detection of silicate emission in the six objects with disks and photospheric silicate absorption feature in one of them without a disk. 

Our main findings are summarized as follows.

\begin{itemize}
    \item By comparing our NIR spectra with standard templates, we find the spectral types of our targets to be between M9.5 and L4 with extinction A$_V$ ranging from 1.3 to 7\,mag. All our targets produce best-fitting results with young templates and for most objects the results are consistent with the literature.
    \item We also compare our NIRSpec spectra with photospheric models and derive effective temperatures between 1600-1900 K. We compare the near-infrared spectra with two dwarfs of similar spectral type, TWA 28 and VHS1256 b, and find significant water absorption features in all our targets. We also see prominent CO$_2$ molecular absorption features in five of our targets, possibly attributable to high metallicity. We note that our targets show diversity in their spectra, especially in the 3-5$\,\mu m$ range, that is not well represented by photospheric models, pointing towards differences in cloud distribution and chemistry as plausible causes.
    \item Six out of our eight targets exhibit mid-infrared excess emission above the photosphere indicating the presence of disks around them. All of them show 10$\,\mu m$ silicate emission features from warm dust in the inner disk. These are the lowest mass objects detected so far with such a feature arising in the disk. We measure the shape and strength of the silicate feature and find it to vary among our targets owing to the differences in grain growth and the degree of crystallization. Analogous to brown dwarfs, most our these FFPMOs show evidence for dust processing and crystallization, demonstrating the potential for the formation of rocky companions in the disks around free-floating planetary-mass objects. 
    \item Additionally, we report the detection of silicate absorption around 10$\,\mu m$ in one of our targets without  a disk. This is the first time silicate absorption arising from clouds in the atmosphere has been found in a young FFPMO. 
    \item We also see molecular emission lines of methane and/or ethylene in a majority of our targets. These features are strongest in CHA1107-7626, as reported in \citet{flagg2025}. Characterization of the molecular emission in these sources requires higher-resolution observations.
\end{itemize}

\needspace{5\baselineskip}

\begin{acknowledgments}
The authors thank the anonymous referee for a constructive report that helped to improve the paper. We also would like to thank Ilaria Pascucci for helpful comments regarding the parameters plotted in Figure 10. BD and AS acknowledge support from the UKRI Science and Technology Facilities Council through grant ST/Y001419/1/. RJ and LF acknowledge support for the JWST-GO-04583.008 program provided by NASA through a grant from the Space Telescope Science Institute, which is operated by the Association of Universities for Research in Astronomy, Inc., under NASA contract NAS 5-03127. V.A-A acknowledges support from the INAF grant 1.05.12.05.03. KM acknowledges support from the Fundação para a Ciência e a Tecnologia (FCT) through the grant 2022.03809.CEECIND, and the Scientific Visitor Programme of the European Southern Observatory (ESO) in Chile. PP acknowledges funding from the UK Research and Innovation (UKRI) under the UK government’s Horizon Europe funding guarantee from ERC (under grant agreement No 101076489). 

\end{acknowledgments}

\vspace{5mm}
\facilities{JWST}

\software{astropy (\citealt{astropy2013,astropy2018})}




\bibliography{main}{}
\bibliographystyle{aasjournal}



\end{document}